\documentclass[12pt]{article}
\usepackage{epsfig, amsmath, amssymb}
\usepackage{graphicx,psfrag}
\usepackage{xcolor}
\newcommand{\nn}{\nonumber}
\newcommand{\be}{\begin{equation}}
\newcommand{\ee}{\end{equation}}
\newcommand{\ben}{\begin{equation}}
\newcommand{\een}{\end{equation}}
\newcommand{\bea}{\begin{eqnarray}}
\newcommand{\eea}{\end{eqnarray}}
\newcommand{\bA}{\begin{array}}
\newcommand{\eA}{\end{array}}
\newcommand{\bc}{\begin{center}}
\newcommand{\ec}{\end{center}}
\newcommand{\al}{\alpha}

\newcommand{\ra}{\rightarrow}
\newcommand{\del}{\partial}

\newcommand{\ie}{{\it i.e.}}
\newcommand{\eg}{{\it e.g.}}

\newcommand{\lan}{\langle}
\newcommand{\ran}{\rangle}

\begin{document}


\begin{titlepage}

%

\bc

\hfill 
\\         [25mm]

{\Huge de Sitter space, extremal surfaces
 \\ [2mm]  and ``time-entanglement''} 
\vspace{16mm}

{\large K.~Narayan} \\
\vspace{3mm}
{\small \it Chennai Mathematical Institute, \\}
{\small \it H1 SIPCOT IT Park, Siruseri 603103, India.\\}

\ec
\vspace{30mm}

\begin{abstract}
  We refine previous investigations on de Sitter space and extremal
  surfaces anchored at the future boundary $I^+$. Since such surfaces
  do not return, they require extra data or boundary conditions in the
  past (interior). In entirely Lorentzian de Sitter spacetime, this
  leads to future-past timelike surfaces stretching between
  $I^\pm$. Apart from an overall $-i$ factor (relative to spacelike
  surfaces in $AdS$) their areas are real and positive. With a
  no-boundary type boundary condition, the top half of these timelike
  surfaces joins with a spacelike part on the hemisphere giving a
  complex-valued area. Motivated by these, we describe two aspects of
  ``time-entanglement'' in simple toy models in quantum mechanics. One
  is based on a future-past thermofield double type state entangling
  timelike separated states, which leads to entirely positive
  structures.  Another is based on the time evolution operator and
  reduced transition amplitudes, which leads to complex-valued
  entropy.
\end{abstract}


\end{titlepage}

{\tiny 
\begin{tableofcontents}
\end{tableofcontents}
}


\vspace{-3mm}

\section{Introduction and summary}

It is of great interest to understand holography for de Sitter space
(see the review \cite{Spradlin:2001pw}).  In de Sitter (and cosmology
more generally) perhaps the natural asymptotics are in the far future
or the far past: this thinking leads to $dS/CFT$
\cite{Strominger:2001pn,Witten:2001kn,Maldacena:2002vr} (and
\cite{Anninos:2011ui} in the higher spin context), which associates a
hypothetical non-unitary dual Euclidean CFT at the future boundary
$I^+$, with several dramatic differences from $AdS$
\cite{Maldacena:1997re,Gubser:1998bc,Witten:1998qj}. A particularly
fascinating question is whether de Sitter entropy
\cite{Gibbons:1977mu} can be understood as some sort of entanglement
entropy. It is then natural to ask if the extensive investigations of
holographic entanglement in $AdS$ \cite{Ryu:2006bv,Ryu:2006ef,HRT}
can be generalized to de Sitter space.

One possible generalization of the Ryu-Takayanagi formulation to de
Sitter space is to consider the bulk analog of setting up entanglement
entropy in the dual Euclidean $CFT$ on the future boundary
\cite{Narayan:2015vda}. We restrict to some boundary Euclidean time
slice as a crutch, define subregions on these slices, and look for
extremal surfaces anchored at $I^+$ dipping into the holographic
(time) direction. Analysing this extremization interestingly shows that
surfaces anchored at $I^+$ do not return to $I^+$, \ie\ there is no
$I^+\ra I^+$ turning point, so there are no spacelike surfaces
connecting points on $I^+$. There exist analytic continuations
of RT surfaces in $AdS$ which lead to complex extremal surfaces
\cite{Narayan:2015vda,Narayan:2015oka,Sato:2015tta,Miyaji:2015yva}. In
\cite{Narayan:2017xca,Narayan:2020nsc}, entirely timelike future-past
extremal surfaces were studied, stretching from $I^+$ to $I^-$.

In this note, we develop this further, stitching together an overall
perspective which hopefully adds value to the understanding of these
studies.  The absence of $I^+\ra I^+$ returns for surfaces implies
that surfaces starting at $I^+$ continue inward, to the past: this
suggests that they require extra data or boundary conditions in the
interior, or far past to be well-defined. One obvious possibility for
an entirely Lorentzian de Sitter space (sec.~\ref{dSfp}) is that the
surfaces then end at the past boundary $I^-$. Analysing this in detail
leads to future-past surfaces stated above
\cite{Narayan:2017xca,Narayan:2020nsc}.  These are timelike extremal
surfaces stretching between subregions at $I^+$ and equivalent ones at
$I^-$: they are akin to rotated analogs of the Hartman-Maldacena
surfaces \cite{Hartman:2013qma} in the eternal $AdS$ black hole. Being
entirely timelike, their area has an overall $-i$ factor, relative to
the familiar spacelike extremal surfaces in $AdS$\ (this overall $-i$
was discarded in \cite{Narayan:2017xca,Narayan:2020nsc}; see below).
Since we obtain codim-2 surfaces (when they exist), their area
scales as de Sitter entropy.

Another possibility for the interior boundary conditions arises from
modifying de Sitter from being entirely Lorentzian in accord with the
Hartle-Hawking no-boundary prescription, \ie\ to cut $dS$ in the
middle and remove the bottom half, replacing it with a hemisphere
(sec.~\ref{dSnb}). Now we join the top timelike part of the extremal
surfaces above with regularity at the mid-slice to a spatial extremal
surface that goes around the hemisphere (thus turning around): see
\cite{Hikida:2022ltr,Hikida:2021ese} for $dS_3$. This spacelike part
has real area so that the total area is complex-valued. The top part
of the surface (in the Lorentzian de Sitter) is the same as in the
entirely timelike surfaces above: this reflects consistency of the
future-past surfaces with Hartle-Hawking boundary conditions. The
finite real part of the area of the no-boundary surfaces arises from
the hemisphere and is precisely half de Sitter entropy for any
dimension when the subregion at $I^+$ becomes maximal.
In sec.~\ref{sec:timeCont-rot}, we give some comments on these
future-past and no-boundary surface areas in terms of time contours,
and argue that they can be regarded as space-time rotations from
timelike subregions in $AdS$-like spaces.

Imaginary values also arise in studies of quantum extremal surfaces in
de Sitter with regard to the future boundary
\cite{Chen:2020tes,Goswami:2021ksw}, stemming from
timelike-separations (sec.~\ref{2dcfttE}). Complex-valued entanglement
entropy was also found quite explicitly in studies of ghost-like
theories, including simple toy quantum-mechanical models of
``ghost-spins'', \eg\ \cite{Narayan:2016xwq,Jatkar:2017jwz}.

For entirely Lorentzian $dS$, the entirely timelike future-past
surfaces are akin to entirely timelike geodesics for ordinary
particles moving in time. Removing the overall $-i$ in their pure
imaginary areas (relative to real spacelike surface areas) is akin to
calling the length of timelike geodesics as ``time'' rather than
``$-i\cdot$space''.
Overall this suggests that the areas of these $dS$ extremal surfaces
with timelike components encode some new object,
``time-entanglement'', distinct from usual spatial entanglement. In
sec.~\ref{timeEE}, we describe two aspects of this in ordinary quantum
mechanics, which incorporate this entry of late and early time
boundary conditions.  One is based on a future-past thermofield-double
state \cite{Narayan:2017xca} (see also
\cite{Arias:2019pzy,Arias:2019zug}) which leads to entirely positive
structures despite the timelike separation. The other involves
the time-evolution operator and ``reduced transition amplitudes'',
giving complex-valued entropy.
As we were preparing this, the work \cite{Doi:2022iyj} appeared with
partial overlap.

\section{$dS$ extremal surfaces from $I^+$,\ boundary conditions}

The simplest place to see the absence of $I^+\ra I^+$ turning points
\cite{Narayan:2015vda} is in the Poincare slicing with planar foliations, so
\be
ds_{d+1}^2 = {R_{dS}^2\over\tau^2} (-d\tau^2 + dy_i^2)
= {R_{dS}^2\over\tau^2} (-d\tau^2 + dw^2 + dx_i^2)\ .
\ee
Here we have singled out $w\in y_i$ as boundary Euclidean time,
without loss of generality. Taking the $w=const$ slice, we consider at
$I^+$ a strip-shaped subregion (the natural subregions consistent with
planar symmetries), with width along $x\in x_i$ and extremal surfaces
anchored from one boundary interface of the strip. This leads to the
area functional and extremization,
\be\label{dSP-surf}
S_{dS} = -i {R_{dS}^{d-1} V_{d-2}\over 4G_{d+1}} 
\int {d\tau\over\tau^{d-1}} \sqrt{1- (\del_\tau x)^2}\quad
\ra\quad (\del_\tau x)^2 = {B^2\tau^{2d-2}\over 1+B^2\tau^{2d-2}}\ .
\ee
where $B^2$ is some constant. The fact that there is a minus
sign relative to the extremization equation in $AdS$ is the reflection
of the absence of turning points back to $I^+$. We see that
$(\del_\tau x)^2\ll 1$ near the boundary $\tau\sim 0$ and remains
bounded with $(\del_\tau x)^2<1$ throughout, for any real $B^2>0$.
(The surfaces with $B^2<0$ are equivalent to analytic continuations
from $AdS$ RT surfaces
\cite{Narayan:2015vda,Narayan:2015oka,Sato:2015tta,Miyaji:2015yva}.)
We will return to this later.

The absence of $I^+\ra I^+$ return implies that the surfaces march
on inward: this suggests they end at $I^-$ if we focus on entirely
Lorentzian de Sitter space. These lead to future-past 
extremal surfaces, timelike codim-2 surfaces stretching from $I^+$
to $I^-$. We describe this now, first in part reviewing the studies in
\cite{Narayan:2017xca,Narayan:2020nsc}. Alternatively we could modify
Lorentzian $dS$ in accord with the Hartle-Hawking no-boundary
prescription replacing the bottom half of $dS$ by a hemisphere, and
then impose a no-boundary type boundary condition on extremal
surfaces. We will discuss these now.

\subsection{Lorentzian $dS$}\label{dSfp}

{\bf Static coordinates:}\ \ These coordinates exhibit static patches
exhibiting time translation symmetry, but allowing analytic extensions
to the entire de Sitter space. We have
\be
ds^2 = -( 1-{r^2\over l^2} ) dt^2
+ {dr^2\over 1-{r^2\over l^2}} + r^2 d\Omega_{d-1}^2\ .
\ee
In the Northern/Southern diamond regions $N/S$, the static patches,
$t$ is time enjoying translation symmetry. Event horizons for
observers in $N/S$ are at $r=l$: the area of these cosmological horizons
is de Sitter entropy. Towards studying the future boundary, we use
$\tau={l\over r} ,\ \ w={t\over l}$, to recast as\
$ds^2 = {l^2\over\tau^2} \big(-{d\tau^2\over 1-\tau^2} + (1-\tau^2) dw^2
+ d\Omega_{d-1}^2 \big)$:
now $\tau$ is bulk time, with $\tau=0$ the future/past boundary
and the future/past universes described by $0\leq\tau<1$. In this case
the boundary at $I^+$ is $R\times S^{d-1}$. We can take the boundary
Euclidean time slice as any $S^{d-1}$ equatorial plane or as the
$w=const$ slice.

\begin{figure}[h] 
\hspace{2pc}
\includegraphics[width=7pc]{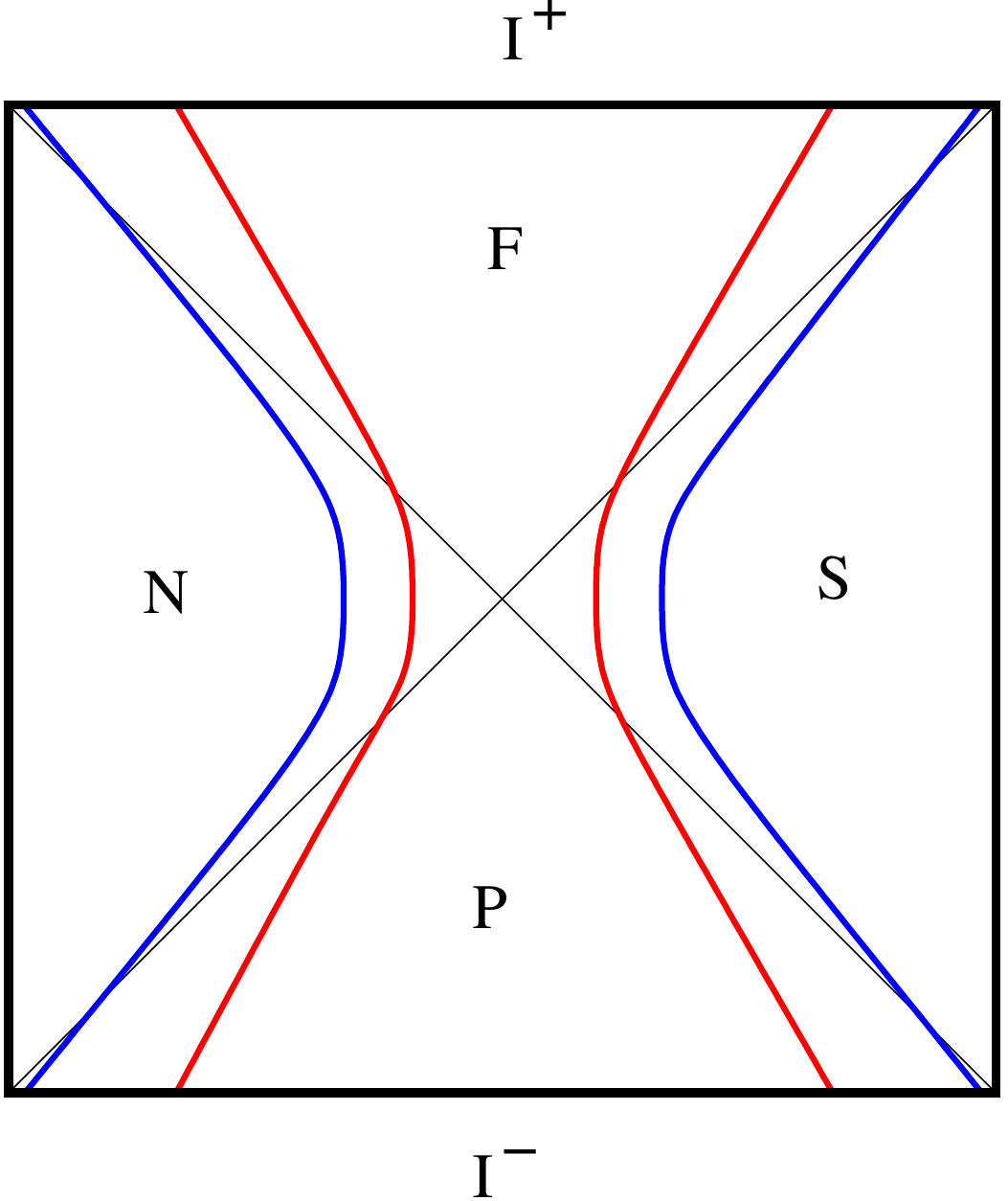}
\hspace{3pc}
\begin{minipage}[b]{24pc}
\caption{{ \label{fig1}
    \footnotesize{$dS$ future-past extremal surfaces stretching
      between $I^\pm$ on an $S^{d-1}$ equatorial plane. The red curve is
      for generic subregion while the blue curve is a limiting curve as
      the subregion becomes the whole space. \newline }}}
\end{minipage}
\end{figure}

Taking the boundary Euclidean time slice as some $S^{d-1}$ equatorial
plane, we define a subregion as $\Delta w\times S^{d-2} \in I^+$ and
an equivalent one at $I^-$. Then we obtain the area functional\
$S=\ -i {l^{d-1} V_{S^{d-2}}\over 4G_{d+1}} \int {d\tau\over\tau^{d-1}}
\sqrt{{1\over f} - f (w')^2}$\
and extremization (with $B^2>0$ some constant)
\be\label{dSs-Surf0}
(1-\tau^2)^2 (w')^2 = {B^2\tau^{2d-2}\over 1-\tau^2
+ B^2\tau^{2d-2}}\,,\qquad
S = -i\,{2 l^{d-1} V_{S^{d-2}}\over 4G_{d+1}} \int_\epsilon^{\tau_*}
{d\tau\over\tau^{d-1}}\ {1\over \sqrt{1-\tau^2 + B^2\tau^{2d-2}}}\,.
\ee
The factor of 2 in the area arises from considering both the top and
bottom parts of the extremal surface (see Figure~\ref{fig1}, reproduced
from \cite{Narayan:2020nsc}).
There is now a real turning point $\tau_*$ at $1-\tau_*^2+B^2\tau_*^{2d-2}=0$
where $|{\dot w}|\ra\infty$: this lies in the $N/S$ diamond regions
where the surface remains timelike. The surface from $I^+$ can be
joined to an equivalent one from $I^-$ (hence the factor of $2$ in $S$
above) which then gives the full, entirely timelike, future-past
surface stretching from $I^+$ to $I^-$. These are rotated analogs of
the Hartman-Maldacena surfaces in the eternal $AdS$ black hole
\cite{Hartman:2013qma}.
There is a limiting surface as $\Delta w\ra\infty$ where the subregion
becomes the whole space $I^\pm$. For $dS_4$ this occurs at $\tau_*=\sqrt{2}$
which corresponds to $B\ra {1\over 2}$\,. 
These surfaces have an area law type divergence (always)
and a finite part: for the limiting surface these are
\be\label{dSs-Surf1}
S^{div} \sim -i {\pi l^2\over G_4} {l\over\epsilon_c}\,,\qquad
S^{fin}\sim -i {\pi l^2\over G_{4}} \Delta w \qquad\qquad [dS_4].
\ee
It is not surprising that we obtain an overall scaling as de Sitter
entropy ${\pi l^2\over G_4}$\,, which is akin to the number of degrees
of freedom in the dual CFT\ (recall that for an $AdS_4$ black hole
the RT surface has area\
$S\sim {R^2\over G_4}\big({V\over\epsilon} + \# T^2 V l\big)$).
These future-past surfaces exhibit various features
\cite{Narayan:2020nsc}: \eg\ the absence of $I^+\ra I^+$ returns
implies that mutual information vanishes.

Considering the $w=const$ slice as the boundary Euclidean time slice,
we consider cap-like subregions defined by $\theta=const$ latitudes
on $S^{d-1}$ at $I^+$ and equivalent ones at $I^-$. Then
\be
S = -i {2l^{d-1} V_{S^{d-2}}\over 4G_{d+1}}
\int {d\tau\over \tau^{d-1}} (\sin\theta)^{d-2}
\sqrt{{1\over 1-\tau^2} - (\theta')^2}
\ee
which is difficult to analyse explicitly for caps at generic $\theta$.
However at $\theta={\pi\over 2}$ it is straightforward to see that
we obtain a future-past extremal surface from the hemispherical cap
on $S^{d-1}\in I^+$ to the corresponding one at $I^-$
\cite{Narayan:2017xca}. This gives area
\be\label{dSs-Surf2}
S = -i {2l^{d-1} V_{S^{d-2}}\over 4G_{d+1}} \int_\epsilon^1
       {d\tau\over\tau^{d-1}} {1\over\sqrt{1-\tau^2}}\ \
       \xrightarrow{\ dS_4\ }\ \ -i {\pi l^2\over G_4} {1\over\epsilon}
\ee
with no finite part.
\begin{figure}[h] 
\hspace{2pc}
\includegraphics[width=6.5pc]{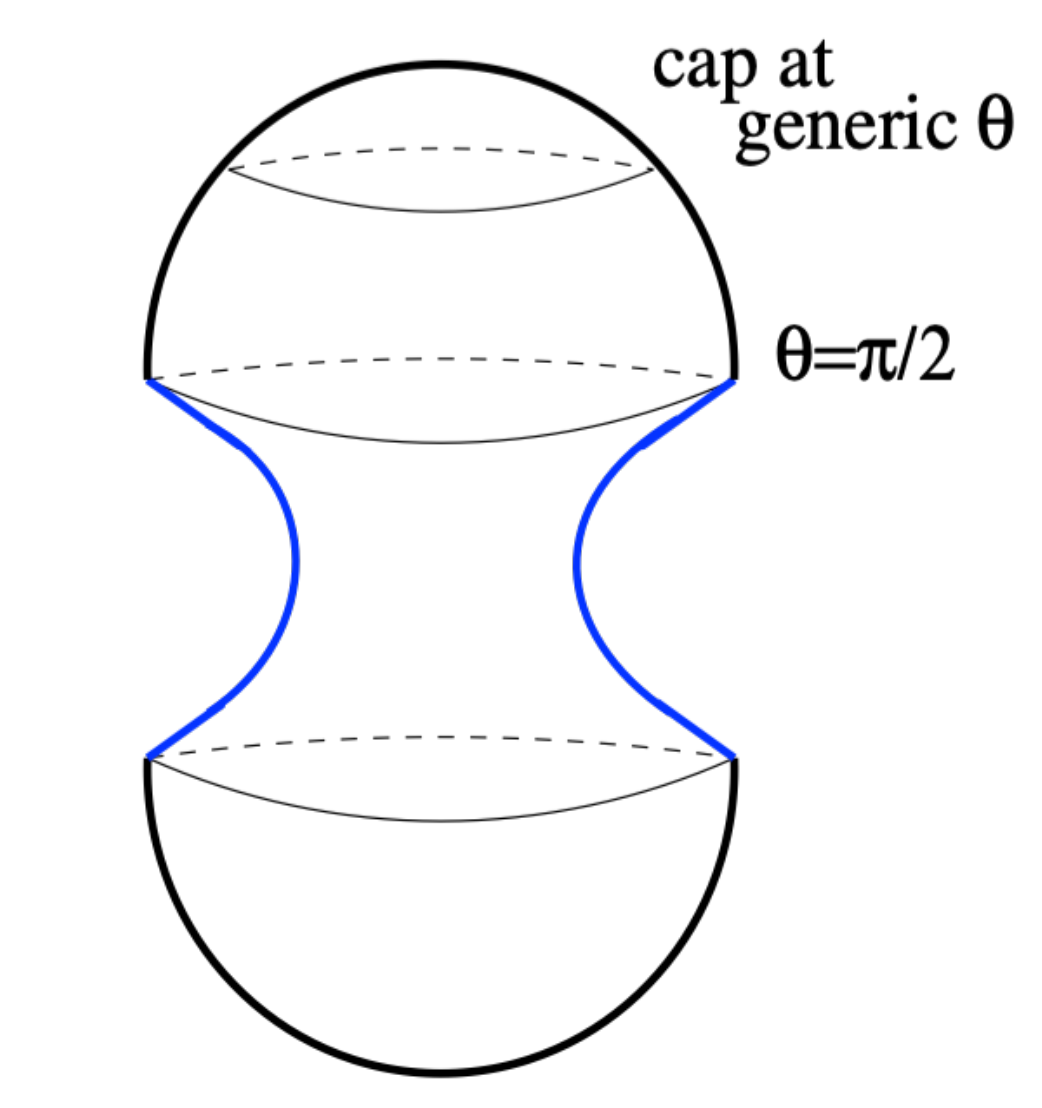}
\hspace{2pc}
\begin{minipage}[b]{26pc}
\caption{{ \label{fig2}
    \footnotesize{Global $dS$ future-past extremal surfaces stretching
      between $I^\pm$ on any $S^d$ equatorial plane in the IR limit
      ($\theta={\pi\over 2}$). This is also the
      picture for the $w=const$ slice in the static coordinates. \newline }}}
\end{minipage}
\end{figure}

\bigskip

{\bf Global:}\ \ Here we have sphere foliations with 
\be
ds_{d+1}^2 = -d\tau^2+l^2\cosh^2{\tau\over l} d\Omega_d^2
\ee
and we can take the boundary Euclidean time slice to be any $S^d$
equatorial plane (which are all equivalent). Then we obtain the area
functional (with factor of $2$ for top+bottom)
\be\label{dSg-Surf}
S = -i {2l^{d-2} V_{S^{d-2}} \over 4G_{d+1}} \int d\tau\ (\cosh\tau)^{d-2}\
(\sin\theta)^{d-2} \sqrt{1-\cosh^2\tau\, (\del_\tau\theta)^2}
\ee
which has structural similarities to the $w=const$ slice above.
At $\theta={\pi\over 2}$ it is straightforward to see a future-past
extremal surface stretching from $I^+$ to $I^-$ with area
(focussing on $dS_4$)
\be\label{dSg-Surfarea}
S = -i {\pi l^2\over G_4} \int_0^{\tau_c/l} d\tau\ \cosh\tau
\sim -i {\pi l^2\over 2G_4} e^{\tau_c/l}
\sim -i {\pi l^2\over 2G_4} {l\over T_c}\ .
\ee
This is an area law divergence type term, with no finite part.
The last expression has been obtained by noting that near $I^+$ we
have\ $ds^2=-d\tau^2+l^2e^{2\tau/l}d\Omega_3^2 \sim
{l^2\over T^2} (-dT^2+l^2d\Omega_3^2)$, with cutoff $T_c=le^{-\tau_c/l}\sim 0$
near $\tau_c\ra\infty$. The area law divergence is structurally
similar to the static coordinates case earlier.

\bigskip

{\bf Poincare:}\ \ The full de Sitter space is obtained from two
Poincare patches joined at the past horizon $\tau\ra-\infty$.  Now
based on the above descriptions for the static and global coordinate
systems, we can likewise construct future-past surfaces by imposing
regularity boundary conditions on the past horizon. For the surface
stretching down from $I^+$ described by the extremization
(\ref{dSP-surf}), we require that the derivatives $\del_\tau x$ match
smoothly onto the corresponding ones for a corresponding surface
stretching up from $I^-$. Note that $(\del_\tau x)^2\ra 1$ as
$\tau\ra-\infty$. The detailed continuation is similar to that
in \cite{Narayan:2017xca,Narayan:2020nsc} for the static coordinates.
This leads to just the area law term again, giving
$S_{dS_4}\sim -i {2l^2\over 4G_4} {V\over \epsilon}$\,.

\subsection{$dS$ no-boundary surfaces}\label{dSnb}

\begin{figure}[h] 
\includegraphics[width=9pc]{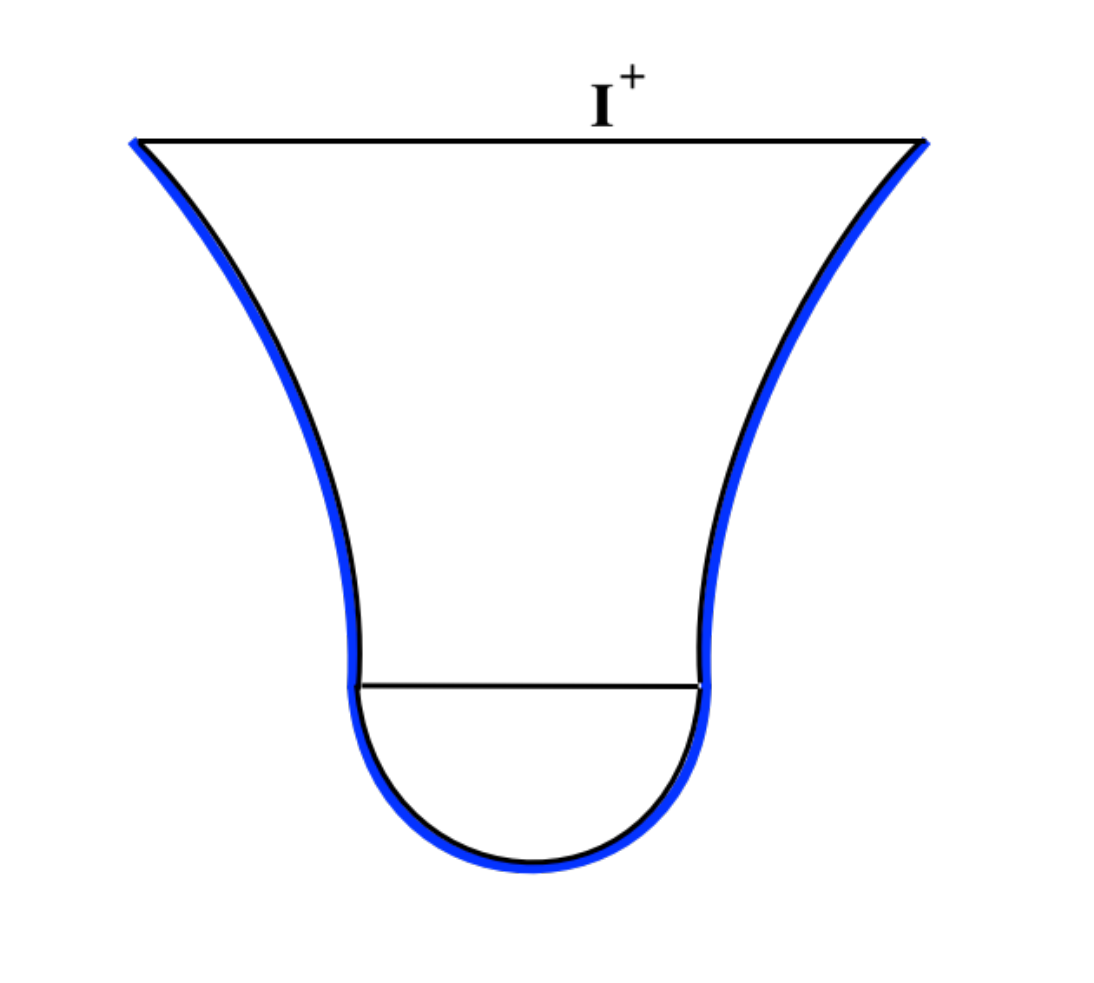}
\hspace{1pc}
\begin{minipage}[b]{28pc}
\caption{{ \label{fig3}
    \footnotesize{Global $dS$ no-boundary extremal surfaces, with a
      top timelike part joining smoothly with a spatial part going
      around the hemisphere in the bottom half. The blue curve is the
      IR limit ($\theta={\pi\over 2}$) on some $S^d$ equatorial plane.
      This is also the picture for the $w=const$ slice in the static
      coordinates. \newline }}}
\end{minipage}
\end{figure}
In accord with the Hartle-Hawking no-boundary prescription
\cite{Hartle:1983ai} (see also \cite{Maldacena:2019cbz}), let us cut
global de Sitter space in the middle, on the $\tau=0$ time slice and
join the top half with a hemisphere in the bottom half: this 
hemisphere is given by the Euclidean continuation
\be\label{dSEuclHemis}
ds^2 = l^2 d\tau_E^2 + l^2 \cos^2\tau_E\,d\Omega_d^2\,;\qquad\quad
\tau=il\tau_E\,,\quad 0\leq\tau_E\leq {\pi\over 2}\,.
\ee
Consider now some $S^d$ equatorial plane (\ie\ $S^{d-1}$) and the
timelike extremal surface in (\ref{dSg-Surf}), at $\theta={\pi\over 2}$
which is the IR limit of such surfaces. The top part of this
surface from $I^+$ hits the $\tau=0$ mid-slice ``vertically'': we join
this smoothly at $\tau=0$ with a surface that goes around the
bottom hemisphere, Figure~\ref{fig3} (see \cite{Hikida:2022ltr} for
$dS_3$)).
This joining being smooth implies consistency with the Hartle-Hawking
prescription.  This IR surface is
\bc
$ds^2 = l^2 d\tau_E^2 + l^2\cos^2\tau_E
(d\theta^2+\sin^2\theta\,d\Omega_{d-2}^2))\Big\vert_{\theta={\pi\over 2}}
= l^2 d\tau_E^2 + l^2\cos^2\tau_E d\Omega_{d-2}^2$\ec
and gives area
\be\label{dSg-hemis}
{l^{d-1}\over 4G_{d+1}} V_{S^{d-2}} \int_0^{\pi/2} d\tau_E\ (\cos\tau_E)^{d-2}
\, =\, {l^{d-1}\over 4G_{d+1}} V_{S^{d-2}}\,
{\sqrt{\pi}\,\Gamma({d-1\over 2})\over 2\,\Gamma({d\over 2})}
\,=\, {1\over 2} {l^{d-1}\,V_{S^{d-1}}\over 4G_{d+1}}\ ,
\ee
using the expression $V_{S^d}={2\pi^{(d+1)/2}\over\Gamma((d+1)/2)}$ for
a $d$-sphere. This real part of the area of this spacelike surface
on the hemisphere is precisely half of de Sitter entropy. This
recovery of the entropy is in detail somewhat different from the
realization of de Sitter entropy as the area of the cosmological
horizon from the point of view of static patch observers. In
particular, one of the hemisphere directions that enters here is
the Euclidean continuation of the time direction in the future universe.

Focussing on $dS_4$, the full area for this no-boundary surface is the
sum of the top timelike part (which is half of the future-past area
(\ref{dSg-Surfarea})) and the hemisphere part becomes
\be\label{SdS4nb}
S = -i {\pi l^2\over 4G_4} {l\over T_c} + {\pi l^2\over 2G_4}\ .
\ee
There are some similarities between these no-boundary surface areas
and the semiclassical Wavefunction $\Psi_{dS}=e^{iS_{cl}}$ for no-boundary
$dS_4$, with $S_{cl}$ the action.  The top Lorentzian half has real
$S_{cl}$ 
which gives a pure phase in
$\Psi_{dS}$. The bottom hemisphere arises after the continuation
(\ref{dSEuclHemis}) to Euclidean time (the no-boundary point is
$\tau_E={\pi\over 2}$ here): $iS_{cl}$ continues to the Euclidean
gravity action\ \ $-\int_{_{_{nbp}}}\sqrt{g}\,(R-2\Lambda)$\ \ pertaining
to the hemisphere, which for $dS_4$ gives\
${1\over 2}\,{l^4V_{S^4}\over 16\pi G_4}\,{6\over l^2}
={\pi l^2\over 2G_4}$\ as is well-known (see \eg\
\cite{Bousso:1995cc,Bousso:1996au}).



A similar calculation of the spatial surface on the hemisphere can be
done for the timelike future-past surface in the static coordinates
discussed earlier. In this case, the boundary was $R_w\times S^{d-1}$
leading to either any $S^{d-1}$ equatorial plane or the $w=const$ slice
as the boundary Euclidean time slice. The Euclidean continuation in
this case is
\be
ds^2 = l^2 (\cos^2\psi\,d\tau_E^2 + d\psi^2 + \sin^2\psi\,d\Omega_{d-1}^2)\,,
\qquad  t=i\tau_E\,,\quad r=l\sin\psi\ ,
\ee
where $\tau_E\in [0,2\pi l]$ and $0\leq\psi\leq {\pi\over 2}$\,.
First, considering the $S^{d-1}$ equatorial plane surfaces, we saw that
there is a limiting surface at $\tau_*>1$ (this is $\tau_*=\sqrt{2}$ for
$dS_4$) which translates to some limiting value $\psi_*$ given by
$\sin\psi_*={r_*\over l}={1\over\tau_*}$\,. Then the surface is
described by
\be
\!ds^2 = \cos^2\psi_* d\tau_E^2+\sin^2\psi_* d\Omega_{d-2}^2
\quad\ra\quad Area = \int_0^{\pi l} \cos\psi_* d\tau_E\, (\sin\psi_*)^{d-2}\,
V_{S^{d-2}}\,{l^{d-2}\over 4G_{d+1}}\,.
\ee
Focussing on $dS_4$ we have $\sin\psi_*={1\over\tau_*}={1\over\sqrt{2}}$
giving the area\
${1\over 2}\,(\pi l)\, {2\pi\,l\over 4G_4} = {\pi^2\,l^2\over 4G_4}$\,.
This apparently unrecognizable value is perhaps not surprising 
due to the limiting surface.

For the $w=const$ slice (equivalently $\tau_E=const$),
the timelike surface from the $\theta={\pi\over 2}$ cap on $S^{d-1}$ leads
on the hemisphere to 
\be
ds^2 = d\psi^2 + \sin\psi^2\,d\Omega_{d-2}^2\quad\ra\quad
Area = {l^{d-1}\,V_{S^{d-2}}\over 4G_{d+1}} \int_0^{\pi/2} d\psi\,(\sin\psi)^{d-2}
= {1\over 2} {l^{d-1}\,V_{S^{d-1}}\over 4G_{d+1}}\ ,
\ee
identical to global $dS$ (\ref{dSg-hemis}) above, not surprising
given the similarities in the calculation for this slice. With the top
timelike part from (\ref{dSs-Surf2}), the total area becomes\
$S = -i {\pi l^2\over 2G_4} {1\over\epsilon} + {\pi l^2\over 2G_4}$\
for $dS_4$.

Note that all these no-boundary surfaces turn around only in the
bottom hemisphere: the top timelike half is identical to the
corresponding future-past surface and there is no $I^+\ra I^+$ turning
point there. Thus if we consider two disjoint subregions the
corresponding no-boundary surfaces are unique (following from the top
halves of the corresponding future-past surfaces), with no new
connected surface emerging: so $S[A\cup B]=S[A]+S[B]$. Thus, as for
the future-past surfaces \cite{Narayan:2020nsc}, mutual information
vanishes here as well.

\subsection{2-dim CFT, timelike subsystems, complex EE}\label{2dcfttE}

Now consider $dS_3$, special for various reasons.
In entirely Lorentzian
global de Sitter, the future-past surfaces on some $S^2$ equatorial plane
slice (\ref{dSg-Surf}) give area\ $S = -i {l\over G_3} \log {l\over T_c}$\,.
If we consider no-boundary $dS_3$, the total area from the top
timelike part and the hemisphere part (\ref{dSg-hemis}) becomes
\be\label{SdS3}
S_{dS_3} = -i {l\over 2G_3} \log {l\over T_c} + {\pi l\over 4G_3}
\ee
The last (real) term is half $dS_3$ entropy ${\pi l\over 2G_3}$\,.
The whole expression can be seen to be an overall $-i$ times the
familiar ${c\over 6}\log {L^2\over \epsilon^2}$\
\cite{Holzhey:1994we,Calabrese:2004eu,Calabrese:2009qy}
with\ $c={3l_{AdS}\over 2G}$ the $AdS_3$ central charge,
alongwith ${l^2\over T_c^2}\ra -{l^2\over T_c^2}$\ (so $\log (-1)=i\pi$).
Note that $dS_3/CFT_2$ has $c_{dS_3}=-i{3l_{dS}\over 2G}$\
\cite{Maldacena:2002vr} which for single intervals would give
imaginary $S$ as for the entirely Lorentzian future-past surfaces
stated above.
So perhaps what is most striking in (\ref{SdS3}) is the real part
arising from the hemisphere which then requires an additional $i$,
which is a novel feature of this Euclidean CFT$_{dS_3}$ dual (in
contrast with ordinary Euclidean CFTs with simply real spatial
lengths and no time). Further related comments appear in 
sec.~\ref{sec:timeCont-rot}.

To put this in perspective, for ordinary unitary 2-dim CFTs, the
entanglement entropy is
\be\label{EE2dcft}
S = {c\over 6} \log {\Delta^2\over\epsilon^2} =
{c\over 6} \log {-(\Delta t)^2+(\Delta x)^2\over\epsilon^2}\ .
\ee
For ordinary spacelike intervals $\Delta^2$, we obtain the familiar
$S = {c\over 3}\log {\Delta x\over\epsilon}$\,. On the other hand
suppose we rotate the subsystem to be entirely timelike with some width
$\Delta t$ in the time direction. This gives
\be\label{ScftTimelike}
S = {c\over 3} \log {\Delta t\over\epsilon} + {c\over 6} (i\pi)
\ee
the imaginary part arising from $\log (-1)$ in the
timelike separation in the interval (more generally the real part
contains $\Delta^2<0$). This imaginary part has appeared
previously in studies of quantum extremal surfaces in de Sitter with
regard to the future boundary \cite{Chen:2020tes,Goswami:2021ksw}.
The bulk matter is modelled as a 2-dim CFT with some central charge
$c>0$ but the timelike separation of the quantum extremal surface
gives $\Delta^2<0$ in (\ref{EE2dcft}) above.

The usual replica formulation of entanglement entropy for a single
interval proceeds by picking the interval $\Delta x\equiv [u,v]$ on
some Euclidean time slice $\tau_E=const$, then constructing $n$
replica copies glued at the interval endpoints. Evaluating
${\rm Tr}\rho_A^n$ can be mapped to the twist operator 2-point
function which then leads finally to the entanglement entropy\
$S_A = -\lim_{n\ra 1} \del_n{\rm Tr}\rho_A^n$. The only data here
is the CFT central charge and the interval in question. The above
Euclidean formulation applies for a timelike interval as well, with
the only change being that the Euclidean time slice is $x=const$ and
the interval is $\Delta t\equiv [u_t,v_t]$. However in continuing back
to Lorentzian time, we rotate $u_t, v_t$, to $-iu_t,\ -iv_t$, and so
we obtain\ $\Delta^2=-(v_t-u_t)^2=-(\Delta t)^2$, which gives
(\ref{ScftTimelike}) above. This of course requires that the CFT 
contains some time direction.

It is also worth noting that complex-valued entanglement entropy
arises quite explicitly in studies of ghost-like theories and simple
quantum mechanical toy models of ``ghost-spins''
\cite{Narayan:2016xwq,Jatkar:2017jwz}: in this case the reduced
density matrix acquires minus signs due to contributions from negative
norm states. Defining contractions over the ghost-spin Hilbert space
appropriately leads to consistent expressions for the reduced density
matrix and entanglement entropy, which are in general complex-valued.

\section{``Time-entanglement'' in quantum mechanics}\label{timeEE}

We have constructed future-past extremal surfaces stretching from $I^+$
to $I^-$. Since they are entirely timelike, their area is pure
imaginary, with an overall $-i$ relative to the area of the familiar
spacelike RT/HRT surfaces in $AdS$. However, apart from this overall $-i$,
the area is real and positive: the overall $-i$ is a uniform factor,
for any subregion at $I^+$. This is a bit reminiscent of the length
of timelike geodesics having an overall $-i$ relative to the length of
spacelike geodesics. We call this timelike length as ``time'' rather
than ``$-i\cdot$space''. This suggests that the areas of the entirely
timelike future-past extremal surfaces encode some new object,
``time-entanglement''.

Recall now the appearance of complex-valued areas for the no-boundary
surfaces which are closely related to the entirely timelike
future-past surfaces: they comprise a timelike component which is
identical to the top half of the future-past one and a spacelike
component from the hemisphere glued in the bottom half. The area is
now complex, with a pure imaginary part from the top timelike
component and a real part from the hemisphere component.

We now describe two aspects of this notion of ``time-entanglement'' in
quantum mechanics (independent of de Sitter at this point). The first
is based on the thermofield-double type state described in
\cite{Narayan:2017xca,Narayan:2020nsc}, while the second is based on
the time-evolution operator, regarding the timelike surfaces as some
sort of transition amplitude.

\subsection{A future-past thermofield double state}

The entirely timelike future-past surfaces, akin to rotated
Hartman-Maldacena surfaces \cite{Hartman:2013qma}, suggest some sort
of entanglement between $I^\pm$, so consider
\be\label{psifpTFD}
|\psi\rangle_{fp} = \sum \psi^{i_n^F,i_n^P} |i_n\ran_F |i_n\ran_P\ ,
\ee
This was written down in \cite{Narayan:2017xca} as an entirely positve
object entangling identical $F$ and $P$ components (with intuition
based on the TFD state for the eternal black hole
\cite{Maldacena:2001kr}). A partial trace
over the second ($P$) copy gives a reduced density matrix with 
nontrivial entanglement entropy. To see how this works, let
us consider a very simple toy example of a 2-state system in ordinary
quantum mechanics. The action of the Hamiltonian $H$ on these
(orthogonal basis) eigenstates and the resulting (simple) time
evolution are
\be\label{qm2state}
H|k\ran = E_k|k\ran\,, \quad k=1,2\ ;\qquad\quad
|k\ran_F \equiv |k(t)\ran = e^{-iE_kt}|k\ran_P\ . \qquad\quad [\lan 1|2\ran=0]
\ee
We consider the $F$ and $P$ slices to be separated by time $t$
and obtain the $F$ state from the $P$ state by time evolution through
$t$.
The future-past TFD state (\ref{psifpTFD}) in this toy case is
\be
|\psi\ran_{fp} = {1\over \sqrt{2}} |1\ran_F |1\ran_P +
{1\over \sqrt{2}} |2\ran_F |2\ran_P
= {1\over \sqrt{2}} e^{-iE_1t} |1\ran_P |1\ran_P
+ {1\over \sqrt{2}} e^{-iE_2t} |2\ran_P |2\ran_P\ ,
\ee
We have normalized the coefficients for maximal entanglement at $t=0$.
For nonzero $t$, there are extra phases due to the time evolution but
they cancel in the reduced density matrix obtained by tracing
$|\psi\ran_{fp}\lan\psi|_{fp}$ over the entire second copy as\
$\delta_{ij}\psi_{fp}^{ki}(\psi_{fp}^*)^{lj}$, so
\be
\rho_{fp} = {\rm Tr}_P |\psi\ran_{fp}\lan\psi|_{fp}
= {1\over 2} |1\ran_F\lan 1|_F + {1\over 2} |2\ran_F\lan 2|_F\ .
\ee
Now imagine a 2-spin analogy, with\ $|1\ran=|++\ran$,\ $|2\ran=|--\ran$,
\ie\ we identify $1\ran, |2\ran$ with the 2-state subspace
$|\pm\pm\ran$ of two spins with states $|\pm\ran$ each for simplicity
and concreteness. Then a partial trace over the second component gives
the reduced density matrix\
$Tr_2\rho_{fp}={1\over 2}|+\ran_F\lan +|_F+{1\over 2}|-\ran_F\lan -|_F$
again with an entirely positive structure, and entropy\ $\log 2$.

If the states in question are not ordinary spins but ``ghost-spins''
with negative norm states, as discussed in \cite{Narayan:2017xca}
based on the studies in \cite{Narayan:2016xwq,Jatkar:2017jwz}, the
fact that we have entangled identical components in both the future
and past copies ensures that the minus signs cancel in
$\gamma_{\sigma\rho}\psi_{fp}^{\al\sigma}(\psi_{fp}^*)^{\beta\rho}$\
(with $\gamma_{ij}$ the indefinite ghost-spin metric) again yielding
an entirely positive structure.

This future-past TFD state with timelike separation is quite different
in principle from the usual TFD state. This positive structure despite
the timelike separation is in some sense similar in spirit to the
areas of the entirely timelike surfaces after stripping off the
universal overall $-i$.

\subsection{Time-evolution and reduced transition amplitudes}

Unlike $AdS$ where specifying boundary data fixes the extremization
problem, $dS$ extremal surfaces starting at late times on $I^+$ do
not return, thus requiring extra data on boundary conditions in the
far past. This is reminiscent of scattering amplitudes, \ie\ final
states from initial states, or equivalently time evolution. It is
then amusing to ask for entanglement-like structures arising from 
the time evolution operator ${\cal U}(t)$ after a partial trace over some
environment: in other words, we look for a ``reduced transition
amplitude'' and its entropy. This suggests (taking $A$ subregion,
$B$ environment)
\be\label{rhot(t)1}
\rho_t(t) \equiv {{\cal U}(t)\over {\rm Tr}\,{\cal U}(0)}
\quad\ra\quad \rho_t^A = tr_B\,\rho_t\quad\ra\quad
S_A = -tr (\rho_t^A\log\rho_t^A)\ .
\ee
The normalization is so we obtain ordinary entanglement structures
at $t=0$, as we will see explicitly.  
To illustrate, consider again the very simple toy example
(\ref{qm2state}) above. Since everything is diagonal here,
the normalized time evolution operator is simple, becoming
\be\label{rhot(t)1-2st}
{\cal U}(t)=e^{-iHt} :\qquad
\rho_t(t) = {1\over 2} e^{-iE_1t} |1\ran_P\lan 1|_P
+ {1\over 2} e^{-iE_2t} |2\ran_P\lan 2|_P
= {1\over 2} |1\ran_F\lan 1|_P + {1\over 2} |2\ran_F\lan 2|_P\ .
\ee
Now recall the 2-spin analogy: $|1\ran=|++\ran$,\ \
$|2\ran=|--\ran$. A partial trace over the second components gives
\be\label{rhotA(t)1-2st}
\rho_t^A = {1\over 2} e^{-iE_1t} |+\ran_P\lan +|_P
+ {1\over 2} e^{-iE_2t} |-\ran_P\lan -|_P\ ,
\ee
\be\label{SA(t)1}
S_A = -\sum_i {1\over 2} e^{-iE_it}\log\Big({1\over 2} e^{-iE_it}\Big)
= {1\over 2} \log 2 (e^{-iE_1t} + e^{-iE_2t})
+ {1\over 2} (iE_1t)e^{-iE_1t} + {1\over 2} (iE_2t)e^{-iE_2t} . 
\ee
Normalizing ${\cal U}(t)$ by its trace at time $t$ gives
${\rm Tr}\rho_t(t)=1$ for all $t$ (not just $t=0$),
modifying (\ref{rhot(t)1})-(\ref{SA(t)1}) to
\bea\label{rhot(t)2}
\rho_t(t) \equiv {{\cal U}(t)\over {\rm Tr}\,{\cal U}(t)}
\quad &\Rightarrow&\quad
\rho_t(t) = \sum_i p_i\, |i\ran_P\lan i|_P\ ,
\qquad p_i={e^{-iE_it}\over \sum_j e^{-iE_jt}}\ , \nn\\
\ra\ \ \
\rho_t^A = \sum_{i} p_i'\, |i'\ran_P\lan i'|_P
\quad &\ra&\quad
S_A = -\sum_i p_i'\,\log p_i'\ ,
\eea
where $H|i\ran=E_i|i\ran$ and the second line arises after partial trace.
There are similarities with pseudo-entropy \cite{Nakata:2020luh}
although the details above look different a priori. 
There are close interrelations between time entanglement above
(entanglement-like structures based on the time evolution operator
regarded as a density operator) and pseudo-entropy: some of these
explorations in quantum mechanics with various interesting new
features appear in \cite{Narayan:2023ebn}, which also elaborates on
some results outlined below.

$\rho_t^A$ resembles an ordinary maximally entangled state at
$t=0$. Any later time $t\neq 0$ gives complex-valued entropy in 
general (although there are real subfamilies: \eg\ (\ref{rhot(t)2})
for the 2-state case contains a single phase $e^{-i(E_2-E_1)t}$ and
gives $S_A$ real).\ Further the different normalizations give different
results in detail, as is already clear in the simple cases above.
Overall these structures resemble the usual finite temperature mixed
state entanglement, except with imaginary temperature, \ie\ $\beta=it$.

There are also related quantities that arise along similar lines.  For
instance the time evolution operator ${\cal U}(t)$ alongwith a
projection operator onto a generic state $|I\ran$ gives\ \ ${\cal
  U}(t)|I\ran\lan I|=|F_I(t)\ran\lan I|$\ where $|F_I(t)\ran$ is the
future state time-evolved from the initial state $|I\ran$. Normalizing
at time $t$ and performing a partial trace gives a reduced transition
matrix which resembles that in pseudo-entropy \cite{Nakata:2020luh}
but with the future state specifically corresponding to the time
evolved state. Relatedly, normalizing at $t=0$ gives different
structures. For instance, projection onto Hamiltonian eigenstates
$|E_I\ran$ and performing partial trace gives simple phases for
$\rho_t^{A,I}$\ (essentially components of (\ref{rhotA(t)1-2st})), so the
corresponding entropy (\ref{SA(t)1}) is of the form \eg\
$iE_It\,e^{-iE_It}$.

\section{Discussion: $dS$ surfaces, time contours, rotations} 
\label{sec:timeCont-rot}

We have seen that the absence of $I^+\ra I^+$ turning points for
$dS$ extremal surfaces anchored at the future boundary leads to either
future-past surfaces or no-boundary surfaces. Since these surfaces are
characterized by area integrals which ultimately reduce to simple
integrals over the time direction, they can be organized and recast in
terms of time contours, which leads to certain clarifications. Towards
this, recall that the future-past and no-boundary surface areas
(\ref{dSg-Surf}), (\ref{dSg-hemis}), are of the schematic form
(with a reduced area functional $a(\tau)$)
\bea
S_{fp} \sim 2\cdot -i S_0 \int_{\tau_{_{cF}}}^{\tau_*} d\tau\, a(\tau)\,,
\qquad\qquad && [\tau: \tau_{_{cF}}\ra\tau_*\ra\tau_{_{cP}}]\,; \nn \\ 
S_{nb} \sim -i S_0 \int_{\tau_{_{cF}}}^{\tau_*} d\tau\, a(\tau)
+ S_0 \int_{\tau_{E*}}^{nbp} d\tau_E\, a_E(\tau_E)\,,\quad &&
[\tau: \tau_{_{cF}}\ra\tau_*\ra nbp]\,,
\eea
where $S_0$ is de Sitter entropy, $\tau_{cF}$ labels the anchoring cutoff
slice at $I^+$ and $\tau_*$ is the bulk point where the surface is going
``vertically down'' (Figures~\ref{fig2}, \ref{fig3}).
$nbp$ refers to the no-boundary point. In the no-boundary surfaces,
the time contour goes along the real time direction
till $\tau_*$ and then along the Euclidean time path till the $nbp$.
As we saw, these simplify in the IR limit to give
\bea\label{Sfpnb-schematic}
&& S_{fp} = -2iS_0\,I[\tau_{_{cF}},\tau_*]\,;\qquad
S_{nb} = -iS_0\,I[\tau_{_{cF}},\tau_*] + {S_0\over 2}\,; \nn\\ [1.5mm] 
&& \qquad\qquad\quad\Rightarrow\qquad  S_{fp}=S_{nb}-S_{nb}^*\ .
\eea
In this light, it is reasonable to think that the future-past
surface is made of two copies of the no-boundary surface, but with
the time contour schematically being\
$[\tau_{_{cF}}\ra\tau_*\ra \tau_{_{cP}}] =
[\tau_{_{cF}}\ra\tau_*\ra nbp]+[nbp\ra\tau_*\ra\tau_{_{cP}}]$.
Then the real parts in the two copies of $S_{nb}$ cancel to give a
pure imaginary $S_{fp}$. Regarding $S_{nb}$ as some time entanglement
entropy arising from one dual boundary Euclidean CFT copy $Z_{CFT}=\Psi_{dS}$
via $dS/CFT$ \cite{Strominger:2001pn,Witten:2001kn,Maldacena:2002vr}
suggests regarding $S_{fp}$ as arising from two copies
$\Psi_{dS}^*\Psi_{dS}$. It would be interesting to flesh this out
more precisely from a replica formulation, perhaps developing
\cite{Lewkowycz:2013nqa} here.

Looking now at the expressions in detail for $dS_3$
and $dS_4$, \ie\ (\ref{SdS3}), (\ref{dSg-Surfarea}),
(\ref{SdS4nb}), we have
\bea
dS_3: &&\quad  S_{fp} = -i\,{l\over G_3}\log {l\over\epsilon}\,;\qquad
S_{nb} = -i\,{l\over 2G_3}\log {l\over\epsilon} + {l\over 2G_3}\,{\pi\over 2}\,.
\nn\\ [1mm]
dS_4: &&\quad S_{fp} = -i\,{\pi l^2\over 2G_4} {l\over\epsilon}\,;\qquad\ \ 
S_{nb} = -i\,{\pi l^2\over 4G_4} {l\over \epsilon} + {\pi l^2\over 2G_4}\,,
\eea
with $\epsilon\equiv T_c$ in the $dS_4$ expressions (\ref{dSg-Surfarea}),
(\ref{SdS4nb}).
For $dS_5$ there are pure imaginary subleading divergent terms as well,
from the timelike $I$ integral in (\ref{Sfpnb-schematic}).\
Writing the $dS_3$ expression as
\be
dS_3: \quad
S_{nb} = -i\,\Big({c\over 3}\log {l\over\epsilon}
+ {c\over 6}\,(i\pi)\, \Big)\,,\qquad\quad c={3l\over 2G_3}\,,
\ee
suggests that these no-boundary surfaces are a rotation from some
surfaces in $AdS_3$, with central charge $c_{AdS_3}=c$\, (recall that
the $dS_3/CFT_2$ central charge is $c_{dS_3}=-ic$): specifically the
overall $-i$ arises from the $AdS_3$ radial integral reinterpreted as
a time integral in $dS_3$. The term inside the brackets is essentially
the entanglement entropy (\ref{ScftTimelike}) for a timelike interval
in 2-dim CFT: the real logarithmic part is a spatial area contribution
in $AdS_3$, while the imaginary part is a timelike contribution). Thus
the real spacelike part of the $dS_3$ surface, from the Euclidean
hemisphere, maps to a pure imaginary, timelike, contribution in $AdS_3$.

The $dS_4$ case (\ref{SdS4nb}) can be similarly recast as
\be
dS_4: \quad S_{nb} = -i\,\Big( {\pi l^2\over 4G_4} {l\over\epsilon} +
i\,{\pi l^2\over 2G_4} \Big)\,,\qquad S_0={\pi l^2\over G_4}\ ,
\ee
which again resembles an overall rotation from an $AdS_4$ surface,
encoded by the overall $-i$. Again, the term inside has a real part
corresponding to half of the Hartman-Maldacena-like spacelike
surface contribution while the imaginary part is a timelike
contribution.
The fact that all de Sitter no-boundary surfaces have area of the
form (\ref{Sfpnb-schematic}), \ie\
\be
S_{nb} = -i\,\Big( S_0\,I + i\,{S_0\over 2} \Big)\,,
\ee
suggests that the surfaces can be regarded as space-time rotations
from timelike subregions in $AdS$-like spaces. In general these are
distinct from analytic continuations of Poincare $AdS$ RT expressions,
which correspond to distinct time contours (along imaginary time paths)
\cite{Narayan:2015vda,Narayan:2015oka,Sato:2015tta}: \eg\ in $dS_4$
those give real negative area. However these can be mapped to other
appropriate analytic continuations from $AdS$ (see \cite{Doi:2022iyj}).

Note that this is consistent with the $dS$ future-past surfaces (see
Figure~\ref{fig1}) being akin to space-time rotations of
Hartman-Maldacena surfaces in the $AdS$ black hole
\cite{Hartman:2013qma}, as discussed in
\cite{Narayan:2017xca,Narayan:2020nsc}. In that case, the $dS$ area
$S_{fp}$ is pure imaginary, with the overall $-i$ encoding the
rotation from a real spacelike area in $AdS$.

The pure imaginary part of the no-boundary $dS_3$ surface area can be
identified with ${c\over 3}\log {l\over\epsilon}$ for a half-size
interval in a Euclidean CFT on a circle \cite{Calabrese:2004eu}: the
future-past surfaces have twice this area, and so correspond to two
copies.  The real spacelike part of the no-boundary area, arising from
a deep interior Euclideanization of de Sitter, presumably indicates
some new IR aspect of the dual Euclidean CFT that encodes ``interior
regularity''.

There are some parallels in the thinking in sec.~\ref{timeEE} via the
time evolution operator and viewing de Sitter space as a collection of
past-future amplitudes \cite{Witten:2001kn}. This suggests using the
S-matrix $|f\ran\lan i|$ with initial and final states appropriate to
$dS$ to analyse entanglement-like structures.  Needless to say, there
are many things to explore here, in quantum mechanics, de Sitter
holography and time.

\vspace{8mm}

{\footnotesize \noindent {\bf Acknowledgements:}\ \ It is a pleasure
  to thank  Abhijit Gadde, Alok Laddha, Shiraz Minwalla and Sandip Trivedi
  for helpful discussions.
  I also thank Tadashi Takayanagi for conversations on the overall $-i$
  following \cite{Narayan:2017xca}, which have influenced my thinking.
  This work is partially supported by a grant to CMI
  from the Infosys Foundation.  }


\end{document}